\documentclass[a4paper,11pt]{article}
\usepackage{jinstpub} 
\usepackage{lineno}
\usepackage{subcaption}
\usepackage{hyperref}
\usepackage{multirow}

\title{Real-time Anomaly Detection for Liquid Argon Time Projection Chambers}

\author[a,*]{S.~Chung,}
\author[a]{J.~Cleeve,}
\author[a]{A.~Malige,}
\author[a]{G.~Karagiorgi,}
\author[b]{L.~Gerlach,}
\author[b]{A.~A.~Pol,}
\author[b]{I.~Ojalvo}
\affiliation[a]{Columbia University,
New York, NY, 10027, USA}
\affiliation[b]{Princeton University,
Princeton, NJ, 08544, USA}

\emailAdd{seokju.chung@columbia.edu}

\abstract{We present a real-time anomaly detection framework for liquid argon time projection chambers (LArTPCs), targeting applications in particle physics experiments such as the Short Baseline Near Detector or the future Deep Underground Neutrino Experiment. These experiments employ detectors that generate and stream high-resolution but sparse images of neutrino and other particle interactions. Our approach utilizes anomaly detection with autoencoders, compressed through knowledge distillation, to enable the detection of anomalous signals in the data through efficient inference on resource-constrained hardware. The framework is targeted for deployment on computing platforms equipped with field-programmable gate arrays, GPUs, or CPUs, allowing low-latency selection of relevant activity directly from the raw detector data stream. We demonstrate that our approach is suitable for the detection and localization of anomalously ``high-multiplicity'' activity, and outline promising applications for LArTPC online data filtering and triggering.}


\arxivnumber{2509.21817} 

\begin{document}
\maketitle
\flushbottom

\section{Introduction}
\label{sec:intro}
Modern detectors for particle physics experiments are capable of generating massive volumes of data, far exceeding storage and processing capabilities for offline analysis. This issue is particularly pressing for next-generation experiments such as the Deep Underground Neutrino Experiment (DUNE)~\cite{Abi_2020}, which aims to operate continually for more than a decade in search of rare signals such as neutrinos from supernova explosions or proton decay. These efforts demand intelligent and highly-efficient data processing systems, capable of selecting relevant data from the detectors in real time and with low latency \cite{bartoldus2022innovationstriggerdataacquisition}.

In this paper, we present a deep learning-based anomaly detection framework as a viable approach to address these challenges. We employ autoencoders to identify anomalous features in detector-generated, high-rate streaming images without relying on labeled datasets or specific physical signal models. This offers the attractive possibility of broadening the physics reach of future liquid argon time projection chamber (LArTPC) experiments beyond the search for predictable or well-modeled rare signals, enabling the search for signals of new physics in a model-agnostic way.

A key challenge is the large size and complexity of such networks, which poses difficulties for real-time deployment on resource-limited hardware. To mitigate this, we use knowledge distillation (KD)~\cite{hinton2015distillingknowledgeneuralnetwork} to compress a high-capacity, unsupervised Teacher autoencoder network into a lightweight, supervised Student network. This Student model retains key performance capabilities while being suitable for execution on resource-constrained hardware.%
\footnote{Project code is publicly available at \url{https://github.com/NevisRAD/RAD4LArTPC}.}

Our approach is detailed in Section~\ref{sec:approach}. Studies of network performance on real LArTPC data are provided in Section~\ref{sec:perf}, followed by a discussion of our findings in Section~\ref{sec:discussion}. Finally, Section~\ref{sec:summary} provides a summary and outline of future promising applications of this approach to LArTPCs.

\section{Approach: Anomaly Detection with Knowledge Distillation}
\label{sec:approach}

The use of autoencoders enables unsupervised anomaly detection by learning a compressed latent representation of the input data. During reconstruction (the sequence of encoding an input image and decoding it into an output image that closely represents the original), statistically uncommon features that deviate from the learned distribution result in higher reconstruction errors. This is quantified in terms of an ``anomaly score''. Higher anomaly score is attributed to inputs that exhibit rare or anomalous features. As the learning process is entirely data-driven, this method does not rely on predefined labels or model-specific expectations, making it highly adaptable and broadly applicable to the search for rare signals in a model-agnostic way.

On the other hand, the resource-intensive nature of autoencoders can limit their applicability on low-power or resource-constrained computational platforms. Following the approach in Ref.~\cite{pol2023kd}, we apply KD to convert an unsupervised Teacher autoencoder into a supervised Student network with significantly reduced computational footprint. During KD, the Student learns to reproduce the Teacher's anomaly scores from the same inputs, enabling similar performance with significantly reduced computational complexity.

Real-time anomaly detection with KD has previously been demonstrated on CMS ECAL data~\cite{CMS-DP-2024-121}. Our work represents the first application of such a framework to LArTPC data. LArTPCs are a widely used detector technology for neutrino experiments, including SBND~\cite{MicroBooNE:2015bmn}, MicroBooNE~\cite{Fleming:2012gvl}, ICARUS~\cite{MicroBooNE:2015bmn}, and DUNE~\cite{Abi_2020}. Their imaging operating principle, high spatial resolution and continuous readout operation make them excellent candidates for anomaly detection.

Our work makes use of the MicroBooNE Open Dataset~\cite{OpenSamples} to evaluate the performance and viability of this approach. We explore the performance of the Teacher and Student networks, investigate their sensitivity to rare features within the data, and demonstrate their feasibility for hardware deployment.

\subsection{Input Data and Preprocessing}
We use publicly available MicroBooNE open samples~\cite{OpenSamples} consisting of simulated neutrino interactions overlaid on real cosmic-ray background interactions collected  by the MicroBooNE LArTPC. While truth labels (neutrino or cosmic) exist, they are not used during training of the Teacher network, which remains unsupervised.

LArTPC detectors such as the one employed by MicroBooNE operate in the following way~\cite{Rubbia:1977zz}: Charged particles produced in neutrino or cosmic-ray interactions inside the LArTPC ionize argon atoms as they pass through the argon volume, creating trails of ionization electrons. Ionization electron trails then drift through the argon, toward a sensor array situated on one side of the detector, with the use of a uniform electric field. The continuous sampling of ionization charge arriving at the sensor array at the edge of the detector produces a continuous stream of data that can be represented in the form of two-dimensional (2D) images of sensor channel (or spatial position) in one dimension, arrival time (or drift position) in the second dimension, and amount of ionization charge as the pixel value. 

For our study, we use exclusively collection wire-sensor data from the MicroBooNE LArTPC. Therefore, each image corresponds to a 2D projection of the ionization activity inside the MicroBooNE LArTPC, and spans $3456$ wire-sensor channels in one dimension, $6400$ 2~MHz-time ticks in the second dimension, and arbitrarily normalized ADC values representing deposited ionization charge as the pixel value. Each image contains the 2D projections of ionization activity corresponding to multiple particle interactions. Different types of particle interactions differ in ionization ``topology'', i.e.~in the way that their resulting ionization activity is distributed within the image. We use these images as inputs to an anomaly detection network, and we seek to identify rare interactions through their anomalous ``topological'' features. 

\begin{figure}[htbp]
    \centering
    \includegraphics[width=\linewidth]{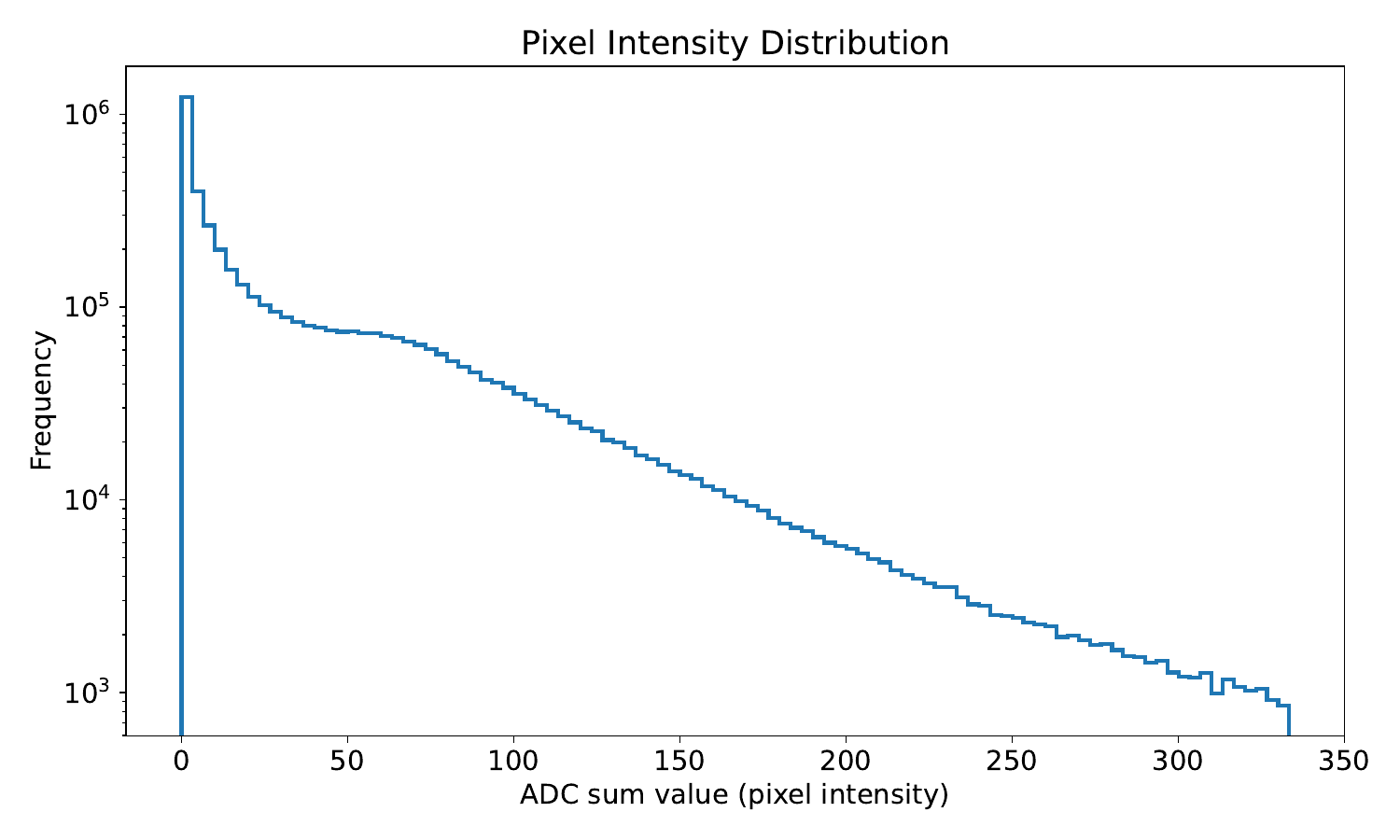}
    \caption{Example of a pixel intensity distribution for a batch of input images after downsampling and before thresholding.}
    \label{fig:pixel_intensity_dist}
\end{figure}
\begin{figure}[htbp]
    \centering
    \includegraphics[width=\textwidth]{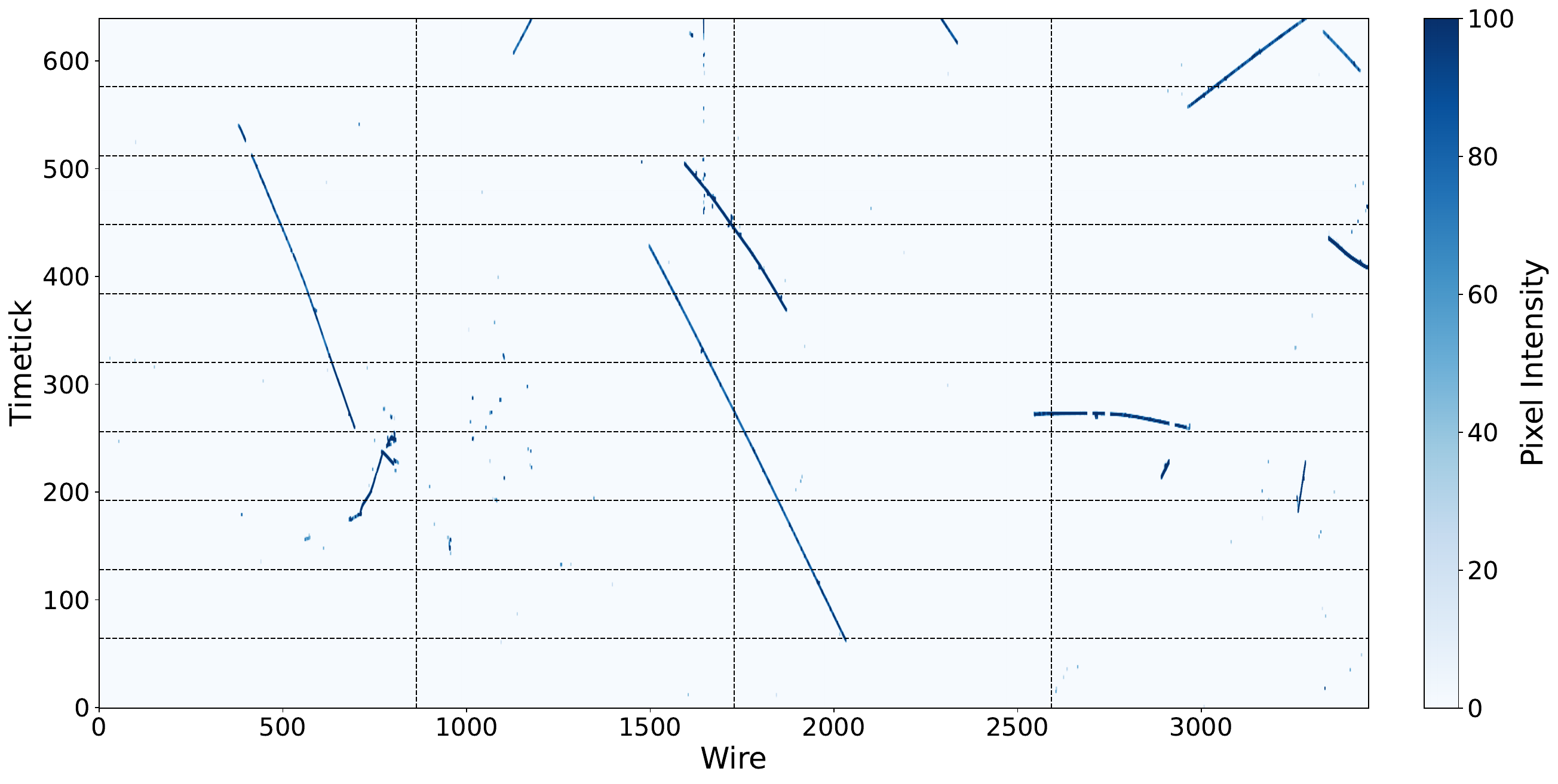}
    \caption{Example of a preprocessed MicroBooNE input image after downsampling and pixel intensity thresholding. The image is divided into smaller segments for network input. Black dotted lines indicate a division into $864 \times 64$-pixel segments. Note that, for readability, the energy–deposition pattern in this figure has been thickened by a $3\times3$ maximum-filter dilation~\cite{VERBEEK1988249}. (This preserves intensity values but visually enlarges non-zero regions to improve readability.) This dilution has not been applied as part of preprocessing.}
    \label{fig:input_image}
\end{figure}

The size of the $3456\times6400$ input image alone would prevent the deployment of the autoencoder model on many commercially available FPGAs, including those in Alveo U250~\cite{AMDUG1289} hardware accelerator cards---a commonly targeted platform. Thus, prior to network input, we downsample the image by a factor of 10 along the time axis. Within each resampling window, the ADC values are summed to define a new \emph{pixel intensity}, which forms the input to the neural network. Pixel intensities below a threshold value of 10 are set to zero, effectively removing noise, while those exceeding a saturation limit of 100 are clipped. These values have been chosen following the visualization example provided from the MicroBooNE Open Dataset documentation. An example of the pixel intensity distribution before cutoff and saturation is given in Figure~\ref{fig:pixel_intensity_dist}. While this potentially limits performance on highly-ionizing interactions, this preprocessing ensures that the input image is both noise-suppressed and bounded in dynamic range, which helps reduce the network computational footprint. Figure~\ref{fig:input_image} shows an example input image after preprocessing. It shows multiple high-ionization depositions identified as ``tracks'' or ``showers''. Those represent the trails of charged particles in different types of interactions. The resulting image, $3456\times640$ pixels in size, is still too large for Alveo U250 deployment. To enable efficient training and deployment, we further subdivide each image into a (variable) number of smaller segments intended as independent input to the network. These smaller inputs allow the model to operate under constrained resource budgets while preserving spatial resolution in the original $3456\times640$ image. 

The total dataset consists of 24,332  $3456\times640$-pixel images, which we split into training, validation, and test sets with respective ratios of 50\%, 10\%, and 40\%. Each image is subdivided into segments; the segment sizes we consider in this work include: $864 \times 64$, $64 \times 32$, and $18 \times 16$. This is illustrated, for example, by the black dotted lines dividing the image in Figure~\ref{fig:input_image} into smaller $864 \times 64$-pixel segments. Each segment is treated as an individual 2D input image for the neural network.

\subsection{Autoencoder-based Anomaly Detection}
Autoencoders are neural networks designed to learn a compressed representation of input data by encoding it into a low-dimensional latent space and reconstructing the original input by decoding the latent space \cite{pmlr-v27-baldi12a}. Inputs that deviate from the learned distribution produce higher reconstruction loss, which we define as the \emph{anomaly score}. This reconstruction loss is calculated as the mean-squared sum over the difference between the pixel values in the input and the output. The autoencoder is trained unsupervised on a given dataset, to minimize this reconstruction loss.



\begin{figure}[htbp]
    \centering
    \begin{subfigure}[b]{0.32\textwidth}
        \centering
        \includegraphics[width=\linewidth]{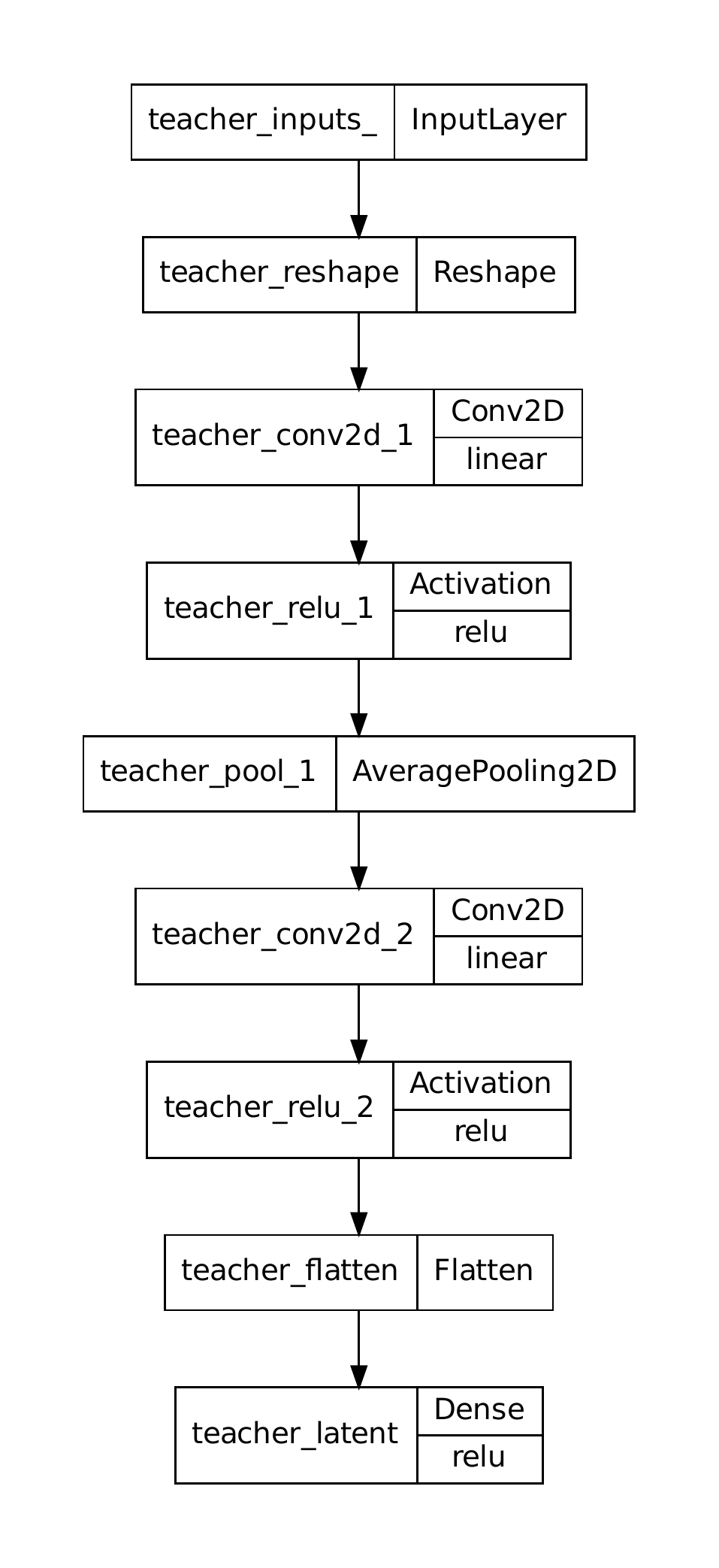}
        \caption{Encoder architecture}
        \label{fig:encoder_architecture}
    \end{subfigure}
    \hfill
    \begin{subfigure}[b]{0.32\textwidth}
        \centering
        \includegraphics[width=\linewidth]{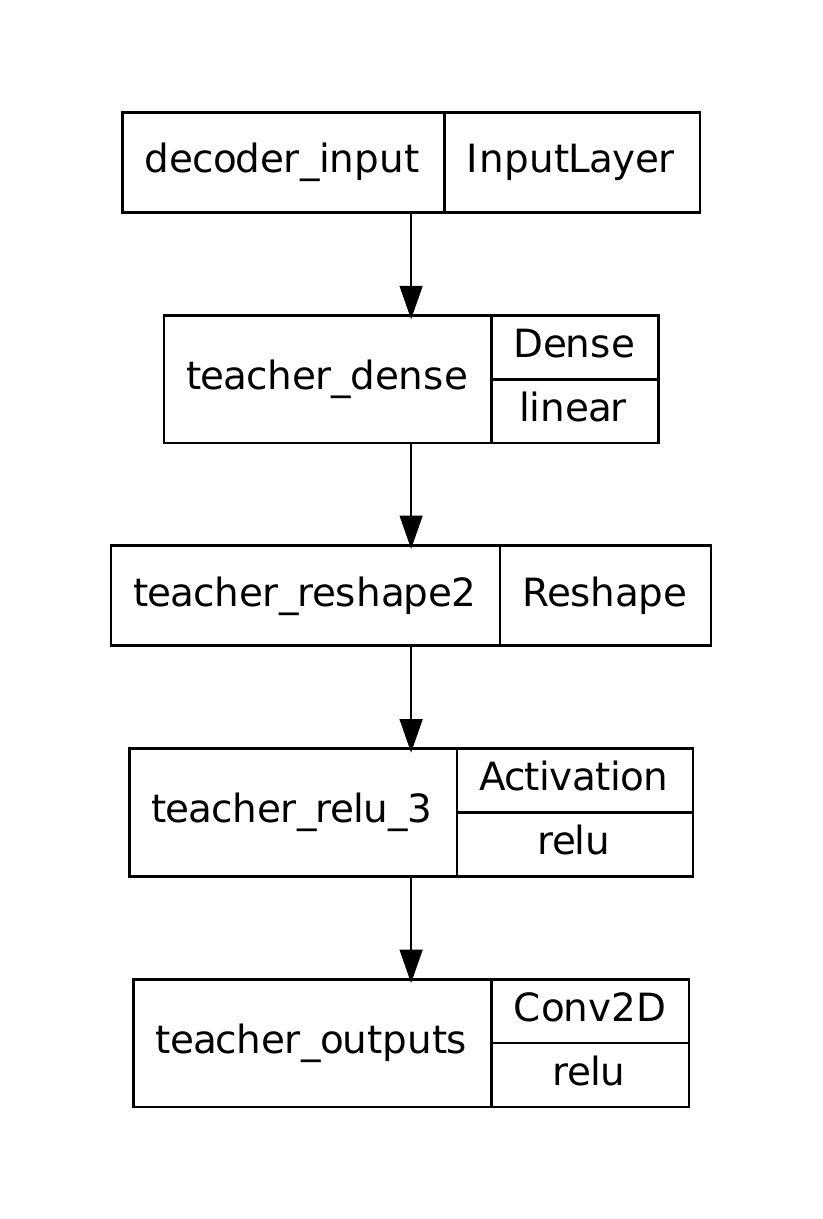}
        \caption{Decoder architecture}
        \label{fig:decoder_architecture}
    \end{subfigure}
    \hfill
    \begin{subfigure}[b]{0.32\textwidth}
        \centering
        \includegraphics[width=\linewidth]{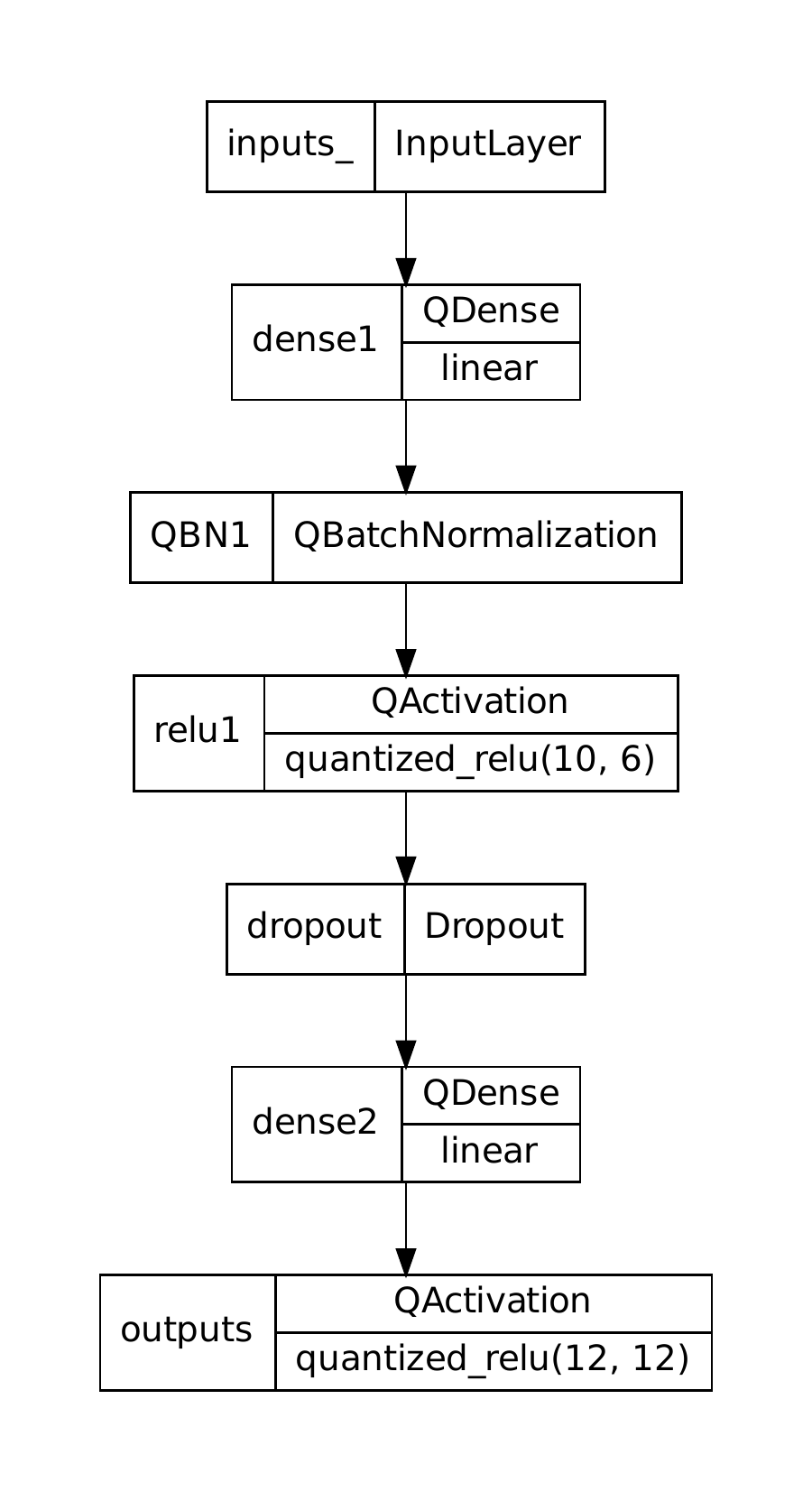}
        \caption{Student architecture}
        \label{fig:student_architecture}
    \end{subfigure}

    \caption{Architectures of the Teacher autoencoder (a, b) and the quantized Student network (c), showing each layer type and activation function, with values used for quantization. Actual layer shapes are scaled corresponding to the input segment size.}
    \label{fig:full_architecture}
\end{figure}

Figures~\ref{fig:encoder_architecture} and~\ref{fig:decoder_architecture} show the architecture of the Teacher autoencoder, where the former (\ref{fig:encoder_architecture}) shows the encoder, and the latter (\ref{fig:decoder_architecture}) shows the decoder. This architecture was inspired from the CICADA autoencoder originally developed for CMS~\cite{CMS-DP-2024-121}, and was adapted in this study for LArTPC inputs. 
Note that the shapes in each layer are scaled corresponding to the input segment shape.

\subsection{Knowledge Distillation for Lightweight Models}
To reduce the computational cost associated with autoencoders, we apply knowledge distillation (KD) to train a compact Student network, following the approach in~\cite{pol2023kd}. Unlike the Teacher, which is an unsupervised autoencoder, the Student is trained in a supervised fashion using the Teacher's anomaly score as the target. This allows the Student to learn to predict anomaly scores directly from inputs, without performing encoding and decoding.

\begin{figure}[t]
    \centering
    \includegraphics[width=\textwidth, trim={0 20 0 0}, clip]{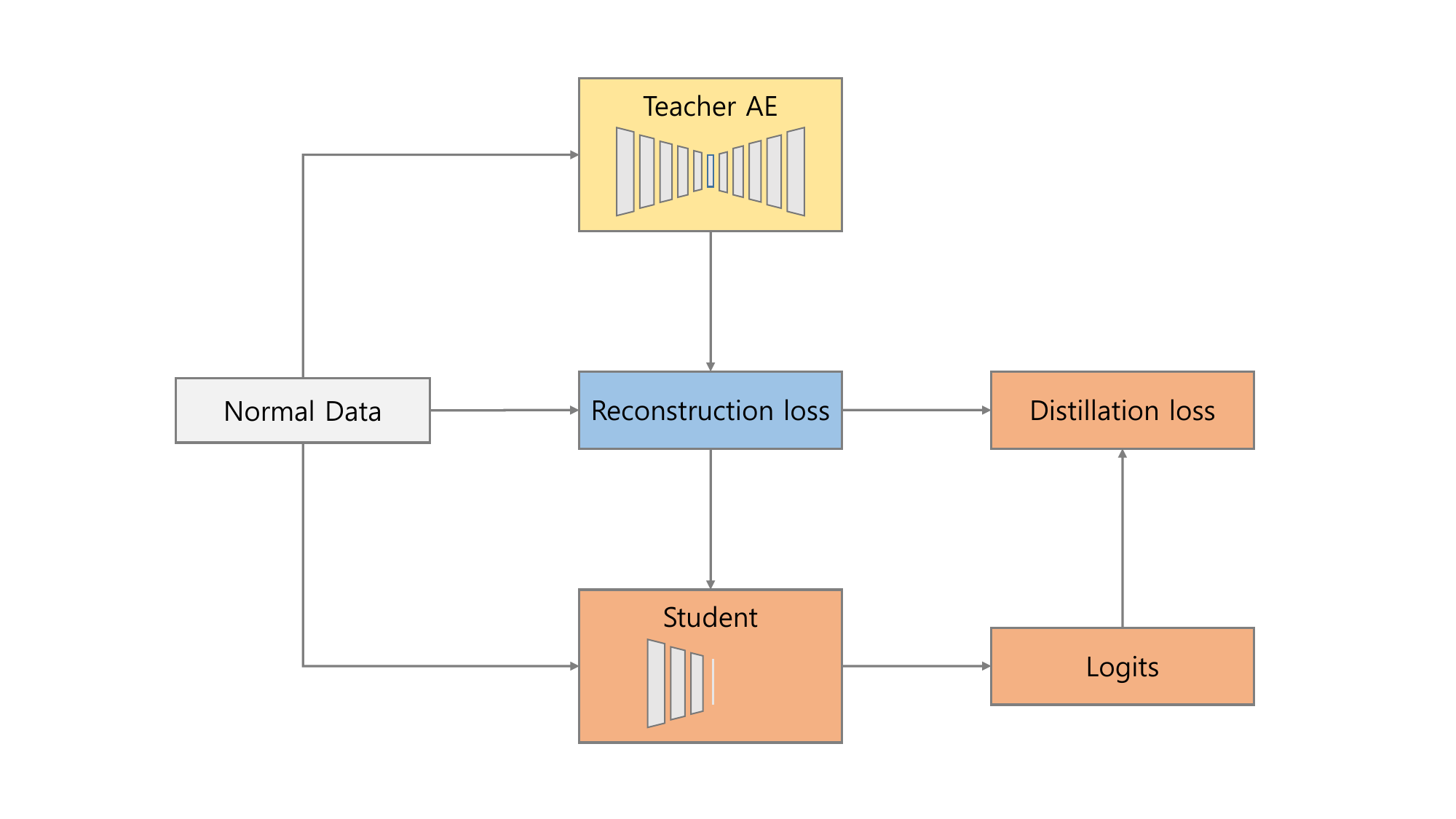}
    \caption{Schematic overview of the training process. The Teacher autoencoder provides anomaly scores that serve as targets for training the Student network~\cite{pol2023kd}.}
    \label{fig:kd_schematic}
\end{figure}

The Student network produces a single scalar output per input, a distilled anomaly score, and is significantly more compact than the Teacher. This compression enables deployment on FPGAs or similar hardware with limited resources. 
The quantized Student architecture is shown in Figure~\ref{fig:student_architecture}. Note that the shapes in each layer are scaled corresponding to the input segment shape. Figure~\ref{fig:kd_schematic} shows a schematic of the entire training process. The Teacher autoencoder is first trained unsupervised for anomaly detection, and its reconstruction loss (anomaly score) is then used as the supervised target for the Student. Quantization-aware training is performed to achieve an efficient conversion to a hardware-deployable version of the Student network.

\section{Network Training, Testing and Evaluation}
\label{sec:perf}

\subsection{Network Training and Validation}

Before evaluating network performance in detail, we first train and validate the expected behavior of both the Teacher and Student models. For the Teacher autoencoder, we expect inputs with rare features to yield high anomaly scores. For the Student, we verify that the knowledge distillation process successfully transfers the Teacher's behavior by examining the correlation between the Teacher and Student anomaly scores. We also calculate the computational footprint reduction achieved through the KD process.

\subsubsection{Teacher Autoencoder}
\begin{figure}[htbp]
    \centering
    \includegraphics[width=\textwidth]{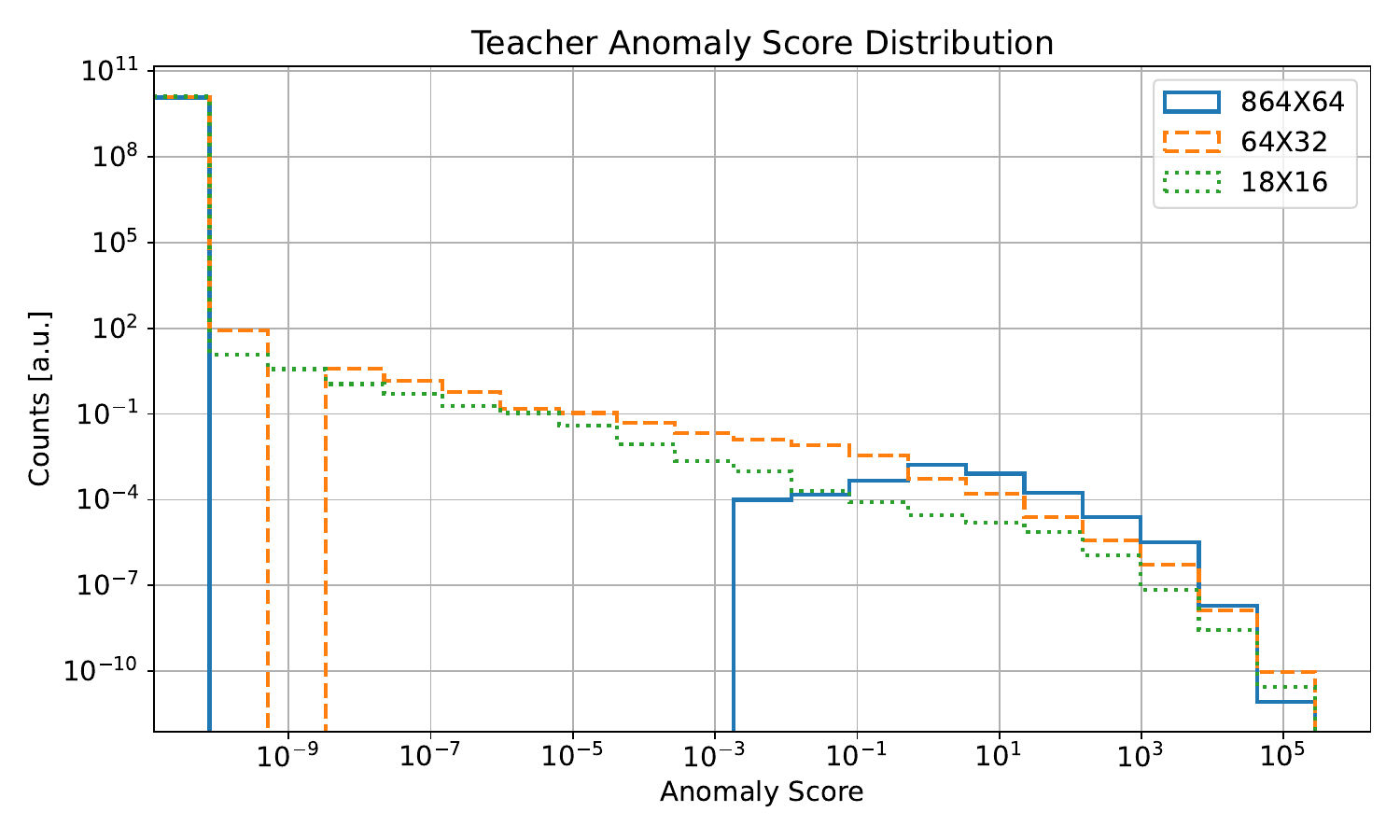}
    \caption{Area-normalized distributions of Teacher anomaly scores across the dataset for different segment sizes. For each of the three cases, high anomaly scores are less frequent, as expected.}
    \label{fig:teacher_histogram}
\end{figure}

Figure~\ref{fig:teacher_histogram} shows the distribution of anomaly scores produced by the Teacher network for different-sized segments. As expected, high anomaly scores are infrequent for all three cases.

\subsubsection{Student Network}
\begin{figure}[htbp]
    \centering
    \includegraphics[width=\textwidth]{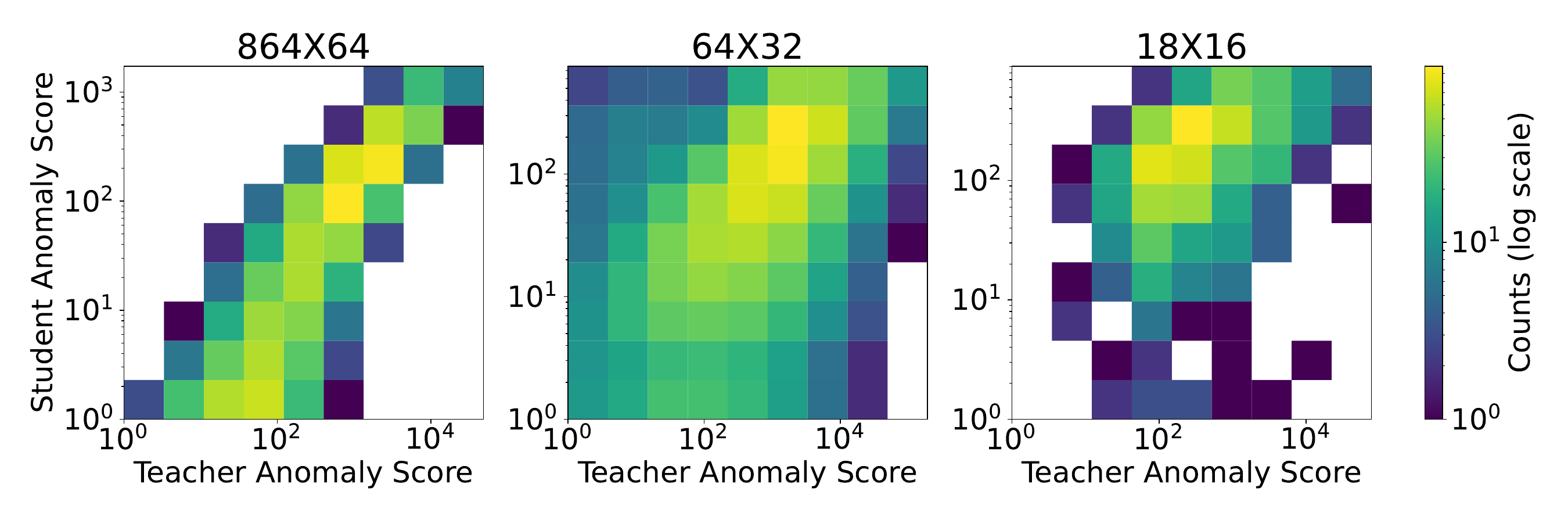}
    \caption{Correlation between Student and Teacher anomaly scores for different segment sizes. Despite being a lightweight model, the Student reasonably approximates the Teacher’s behavior. Anomaly scores below 1 are excluded from these distributions.}
    \label{fig:student_vs_teacher}
\end{figure}

Figure~\ref{fig:student_vs_teacher} shows the correlation between Teacher and Student anomaly scores for different-sized input segment sizes. Note that the range of Student scores is generally narrower than that of the Teacher due to quantization and precision limitation, applied for hardware efficiency. Overall, the Student reproduces the Teacher’s outputs with a consistently positive correlation, although the degree of correlation reduces with decreasing segment size. 

Table~\ref{tab:kd_reduction_factor} shows the number of parameters for the Teacher and Student models, along with the slope and $R^2$ of a linear fit ($y=mx+b$) between their anomaly scores. Overall, the Student achieves a parameter reduction of more than 75, with the smallest Student model being only $\sim100~\mathrm{kB}$ in size.

\begin{table}[htbp]
    \centering
    \caption{Comparison of Teacher and Student model number of parameters, reduction factors,  linear fit slope, and correlation coefficient ($r=\sqrt{R^2}$) of anomaly scores for different input segment sizes.}
    \label{tab:kd_reduction_factor}
    \begin{tabular}{c|cccccc}
    \hline
    {Segment Size} & {Teacher} & {Student} & {Reduction Factor} & {Slope} & {Offset} & {$r$}\\
    \hline
    864 $\times$ 64 & 66,789,361 & 884,833 & 75.5 & $8.54$ & -2.81 & 0.94 \\
    64 $\times$ 32 & 2,492,401 & 32,865 & 75.9 & $17.7$ & 3.02 & 0.51\\
    18 $\times$ 16 & 367,201 & 4,705 & 78.0 & $8.69$ & 0.01 & 0.42\\
    \hline
    \end{tabular}
\end{table}

\subsubsection{Visual Inspection of Anomalous Segments}
\begin{figure}[htbp]
    \centering
    \includegraphics[width=\textwidth]{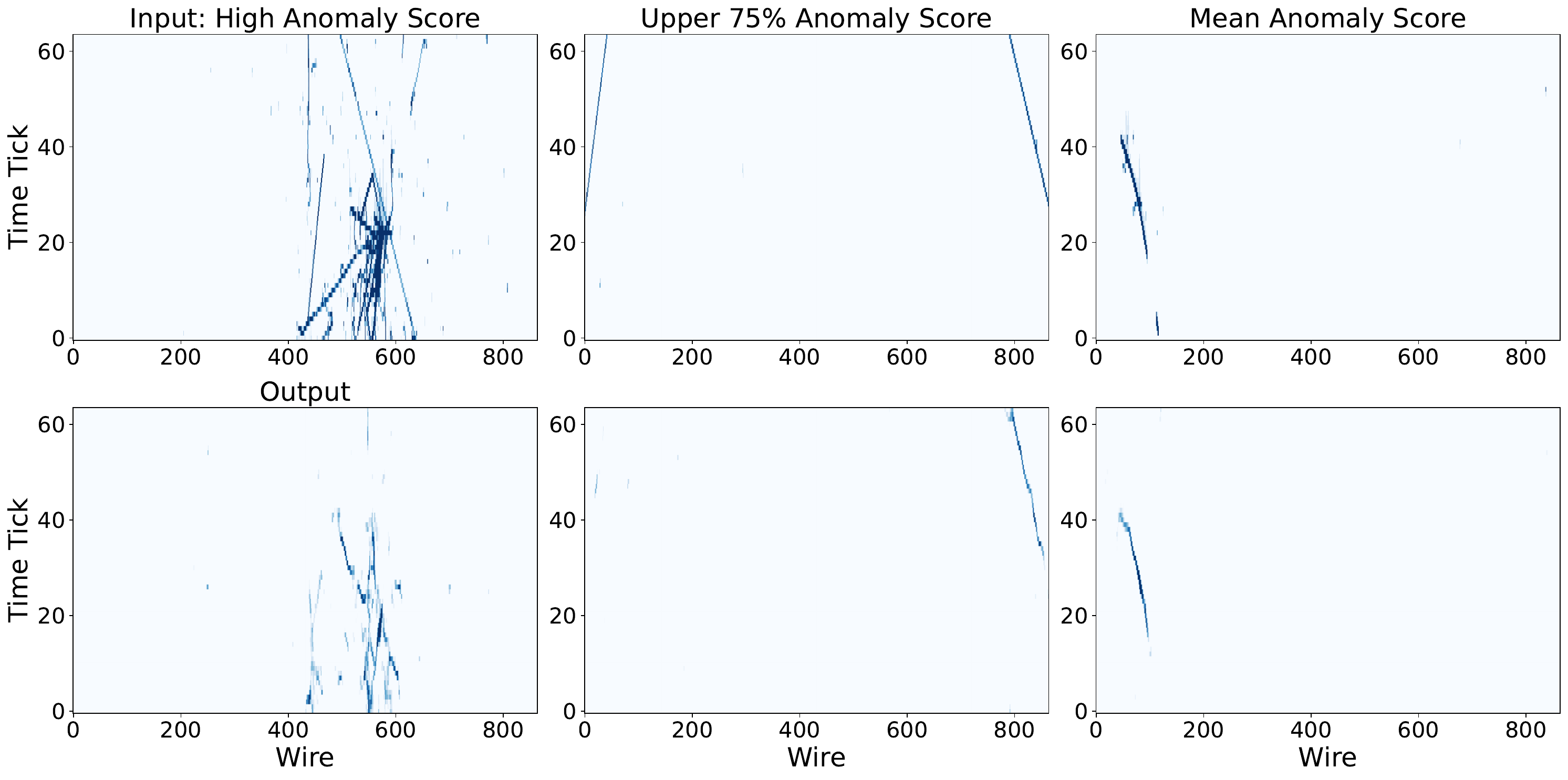}
    \caption{Examples of Teacher inputs (top) and reconstructed outputs (bottom). From a random batch, the input segment with the highest (left), the upper 75\% (middle), and the mean (right) anomaly score was selected. 
    Multi-track segments exhibit large discrepancies, resulting in high anomaly scores. The color scale is the same as in Figure~\ref{fig:input_image}.}
    \label{fig:qualitative_anomaly}
\end{figure}

To understand what types of segments receive high anomaly scores, we visually inspected and compared $O(100)$ input-output pairs with high and low anomaly scores from the Teacher network. Figure~\ref{fig:qualitative_anomaly} presents representative examples of these segment pairs, where inputs with multiple tracks show large reconstruction differences relative to the output, and correspondingly high anomaly scores.

This behavior is expected: the Teacher autoencoder learns the common structure of the training dataset, which, for the MicroBooNE dataset, is dominated by either empty segments or segments with only a single track. Inputs with topology that deviates from this learned distribution, such as those with multiple tracks, or ``high multiplicity,'' are harder to reconstruct, resulting in higher anomaly scores.

\subsubsection{Correlation with Number of Tracks}

To validate the hypothesis that high anomaly scores correspond to multi-track segments, we compare the anomaly score with a track-count metric derived from simulation truth-level information. The MicroBooNE dataset consists of simulated beam neutrino interactions (which contain truth-level labels) overlaid with beam-off data collected by the detector (which is unlabeled data). Because this validation test relies on truth-level information, segments with only simulated beam neutrino interactions were used.

Each simulated image in the MicroBooNE dataset includes a table containing Monte Carlo truth-level information about individual particle trajectories (\texttt{particle\_table/g4\_id})~\cite{OpenSamples}. This identifier distinguishes between different Geant4~\cite{GEANT4:2002zbu}-simulated particles in a single interaction. Image-level information is stored in the signal-level table (\texttt{hit\_table}), in the form of ``hits'', which record signal activity (proportional to pixel intensity) on individual wires and time ticks (which can be mapped into pixels) as a result of ionization charge depositions.

\begin{figure}[t]
    \centering
    \includegraphics[width=\textwidth]{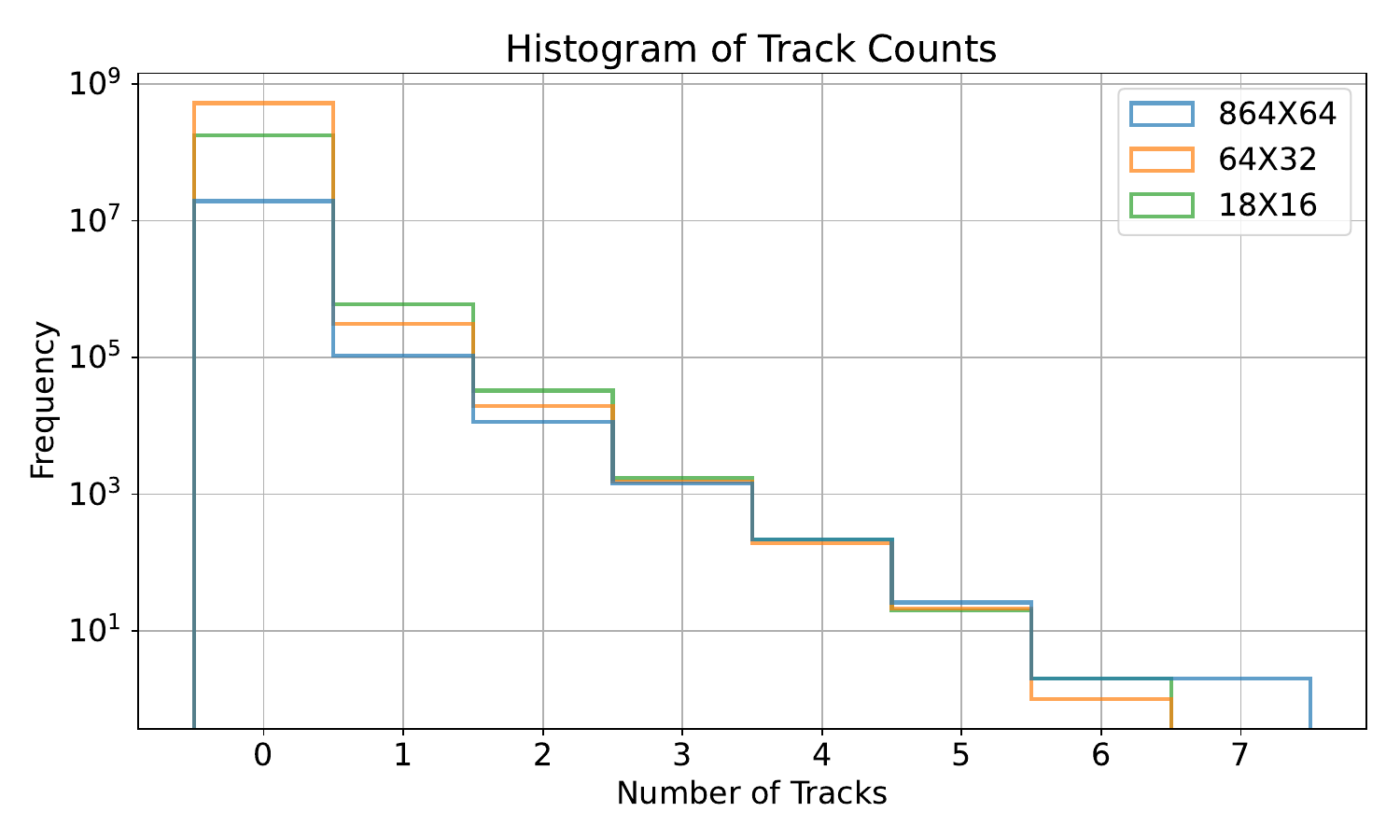}
    \caption{Distribution of truth-level track counts per segment. As expected, segments with high multiplicity are less frequent.}
    \label{fig:track_distribution}
\end{figure}

To construct a segment-level track metric, we associate each hit to a unique track based on its Geant4 identifier, and spatially cluster the hits  before we can count the number of distinct tracks. This clustering accounts for short-range overlaps and is robust to small fluctuations in reconstruction. For simplicity, we define the number of tracks $N_{\mathrm{track}}$ per segment as the number of unique Geant4 id's associated with hit clusters in that region. For each cluster to be counted as unique, it needs to have at least five different hit instances.

\begin{figure}[t]
    \centering
    \includegraphics[width=\textwidth]{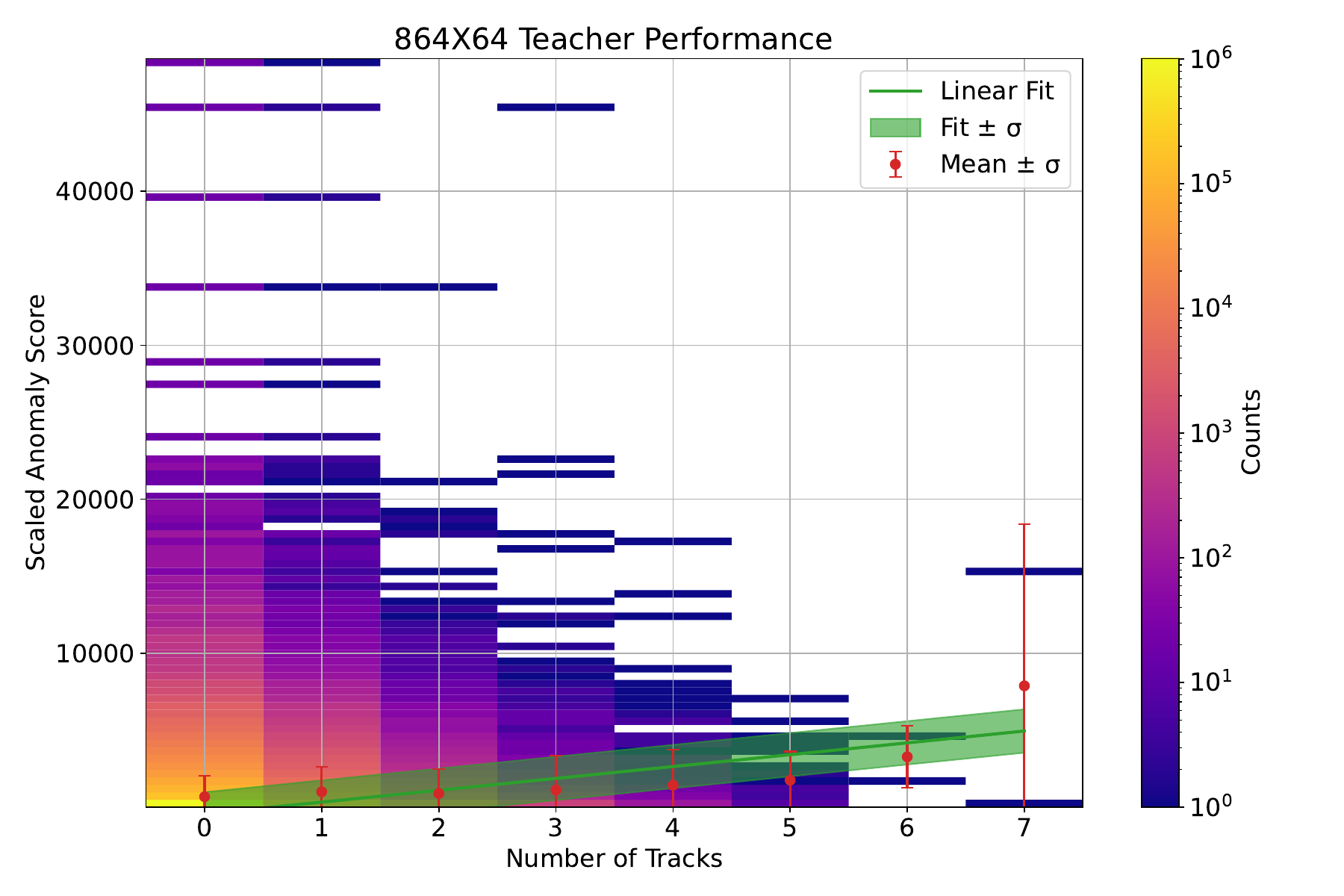}
    \caption{Correlation between Teacher anomaly score and the number of truth-level tracks $N_{\mathrm{track}}$ per segment, derived from Monte Carlo particle and hit association. Segments with multiple tracks yield higher anomaly scores, confirming the model’s sensitivity to complex topologies. The data points are the mean of the anomaly score, the error bars correspond to the standard deviation. The zero-anomaly-scores have been removed from the figure and the fit.}
    \label{fig:raw_score}
\end{figure}

Figure~\ref{fig:track_distribution} shows the distribution of truth-level track counts per segment, confirming the qualitative expectation that multi-track segments are comparatively rare. Figure~\ref{fig:raw_score} illustrates the correlation between the Teacher anomaly score and the number of truth-level tracks, $N_{\mathrm{track}}$, across all evaluated segments. The observed monotonic increase in the average anomaly score with $N_{\mathrm{track}}$ demonstrates the network’s sensitivity to the number of distinct ionizing particles in an input. To assess whether the observed increase of anomaly score with $N_{\mathrm{track}}$ is statistically distinguishable from zero, we performed a linear regression using the \texttt{scipy.stats.linregress} module~\cite{2020SciPy-NMeth}, which tests the null hypothesis of zero slope. The resulting $p$-value for Figure~\ref{fig:raw_score} is $1.5\times10^{-2}$. Owing to the large number of segments in the sample, even a small positive slope is found to be statistically significant. This significance should be interpreted only as evidence of a positive correlation, not as an indication that the anomaly score provides powerful discrimination between different track multiplicities.

\begin{figure}[h]
    \centering
    \includegraphics[width=\textwidth]{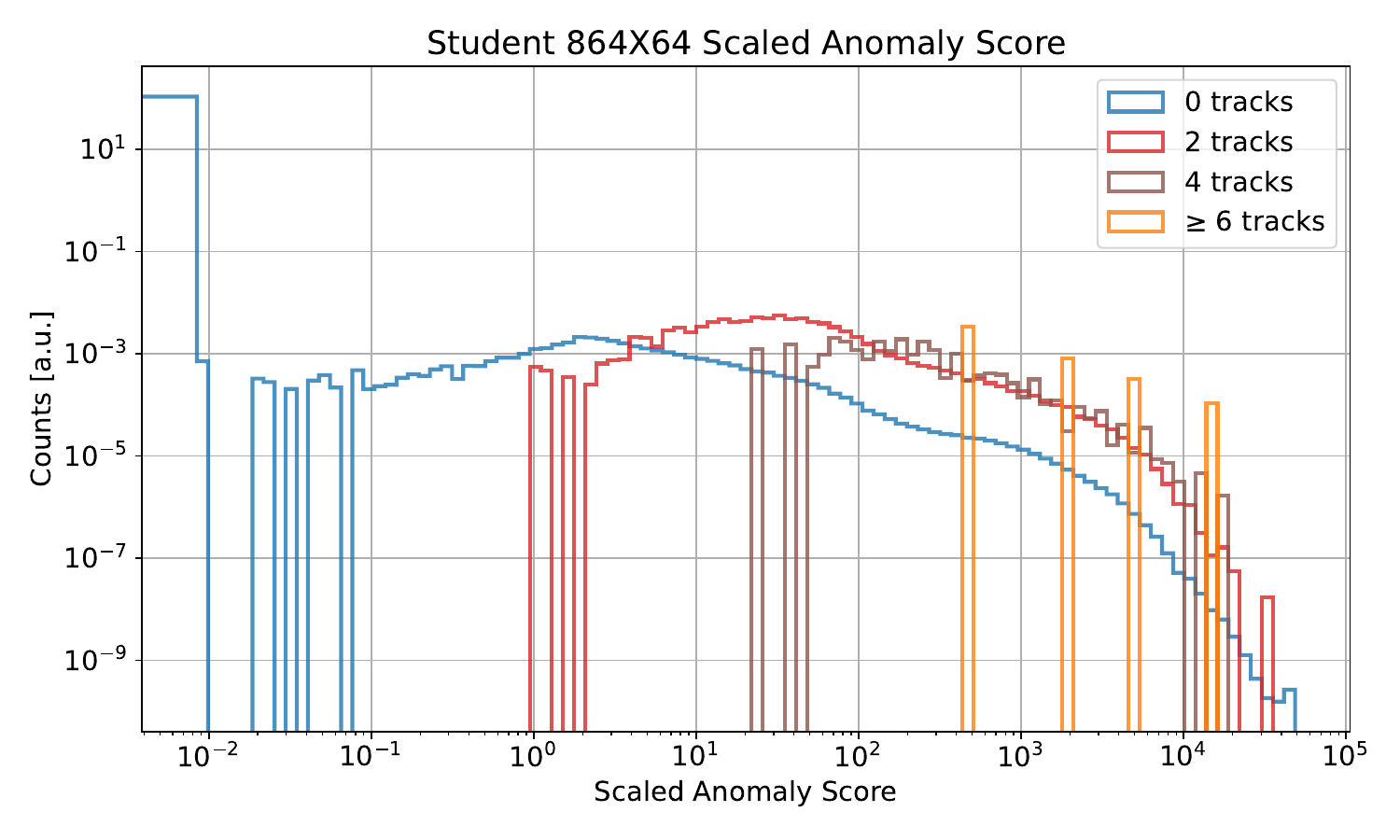}
    \caption{Area-normalized distributions of Teacher anomaly score for segments with 0, 2, 4, and $\geq$6 tracks, using the $864\times64$ Teacher network. The empty segments are sharply peaked near zero, while multi-track segments populate the higher end of the score spectrum, demonstrating the network’s ability to distinguish topologically complex interactions.}
    \label{fig:teacher_distribution}
\end{figure}

Figure~\ref{fig:teacher_distribution} shows the distribution of the Teacher anomaly score for different input classes: empty segments containing zero truth-level tracks, and segments containing two, four, and six or more tracks. The multi-track density has been arbitrarily scaled to account for the much greater abundance of empty compared to multi-track segments. The empty segment distribution is sharply peaked at low anomaly scores, indicating that the autoencoder successfully reconstructs the more common topologies of no or simple interactions (e.g.~a through-going cosmic ray muon). In contrast, high-multiplicity segments exhibit a broader distribution with a long tail toward higher anomaly scores. This reflects the rarity of these topologies and the difficulty of the model to learn the increased complexity and variation present in multi-track topologies. This separation highlights the model’s capacity to act as a low-level data selector by identifying complex or anomalous activity directly from raw wire data.

\begin{figure}[htbp]
    \centering
    \includegraphics[width=\textwidth]{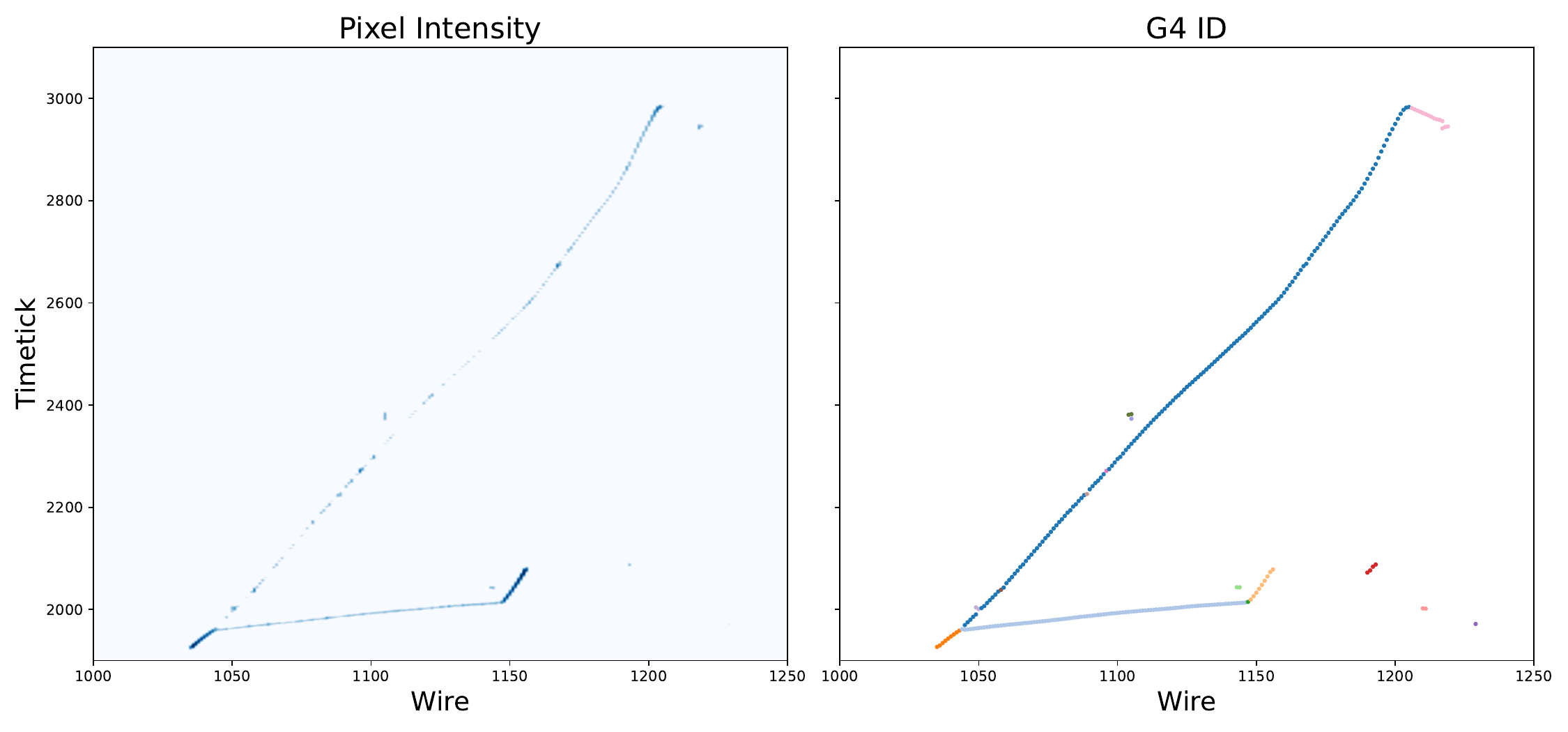}
    \caption{Visual comparison between a pixel-intensity input (left) and its corresponding truth-level track overlay (right). The input exhibits multiple ionization trails, each corresponding to a set of multiple pixels with same \texttt{g4\_id} value. Note that the decay electron (pink color in upper right) track is missing in the wire input due to the threshold we impose as part of prepossessing.}
    \label{fig:image_vs_truth}
\end{figure}

In addition to this numerical correlation, we also visually confirmed that pixel-intensity inputs match well with the corresponding truth-level particle information. Figure~\ref{fig:image_vs_truth} shows an example where a high-anomaly score input aligns with a truth overlay containing multiple overlapping tracks. This reinforces the interpretation that the model assigns higher anomaly scores to rare, high-complexity events (interactions).

\subsubsection{Scaled Anomaly Score Correlation}

Because the anomaly score is defined as a reconstruction loss that measures pixel intensity differences between the input and output segments, it is expected that the anomaly score should also be positively correlated with the sum of pixel intensity. Since the loss is computed as the sum of the squares of pixel-wise differences, segments with higher total pixel intensity naturally yield larger absolute differences when reconstruction errors occur, leading to larger potential deviations. Therefore, higher-intensity inputs tend to produce higher anomaly scores. 

\begin{figure}[t]
    \centering
    \includegraphics[width=\textwidth]{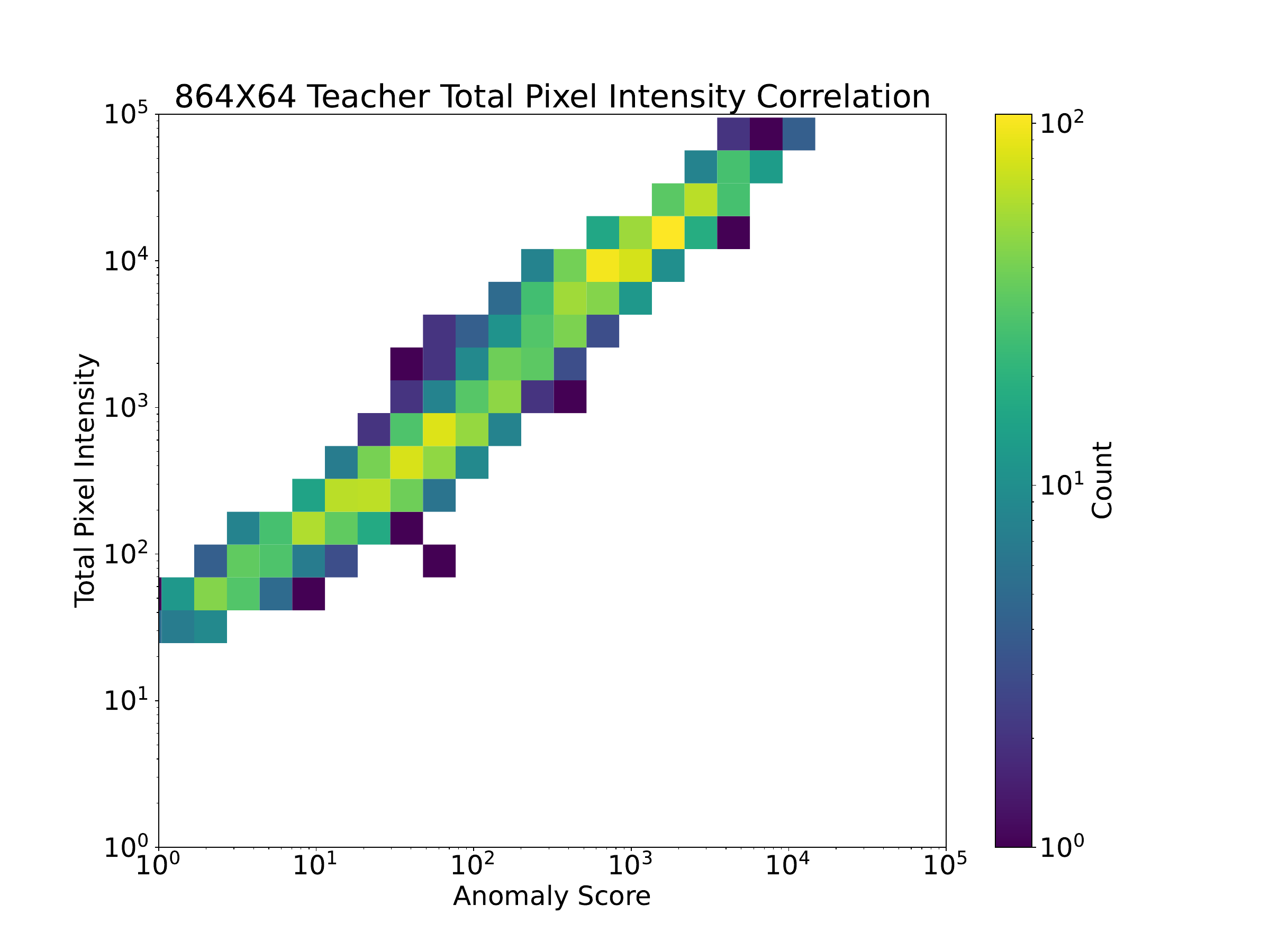}
    \caption{Correlation between Teacher anomaly score and total pixel intensity. A positive trend indicates that segments with greater ionization activity---and therefore larger possible contributions to reconstruction loss---are harder to reconstruct accurately. Anomaly scores below 1 are excluded from these distributions.}
    \label{fig:teacher_correlation}
\end{figure}

Figure~\ref{fig:teacher_correlation} displays the correlation between the Teacher anomaly score and the total pixel intensity of each input segment, for $864\times64$-sized input segments. A clear positive correlation is observed, consistent with the understanding that segments with higher pixel intensity have a larger reconstruction burden, and therefore exhibit higher loss values.

Since anomaly scores are trivially correlated with total pixel intensity, we divide each anomaly score by the total pixel intensity of the corresponding segment, defining this ratio as our scaled anomaly score. This enables us to isolate the model's sensitivity to geometric complexity (e.g., number of tracks) independently of signal strength.

\begin{figure}[t]
    \centering
    \includegraphics[width=\textwidth]{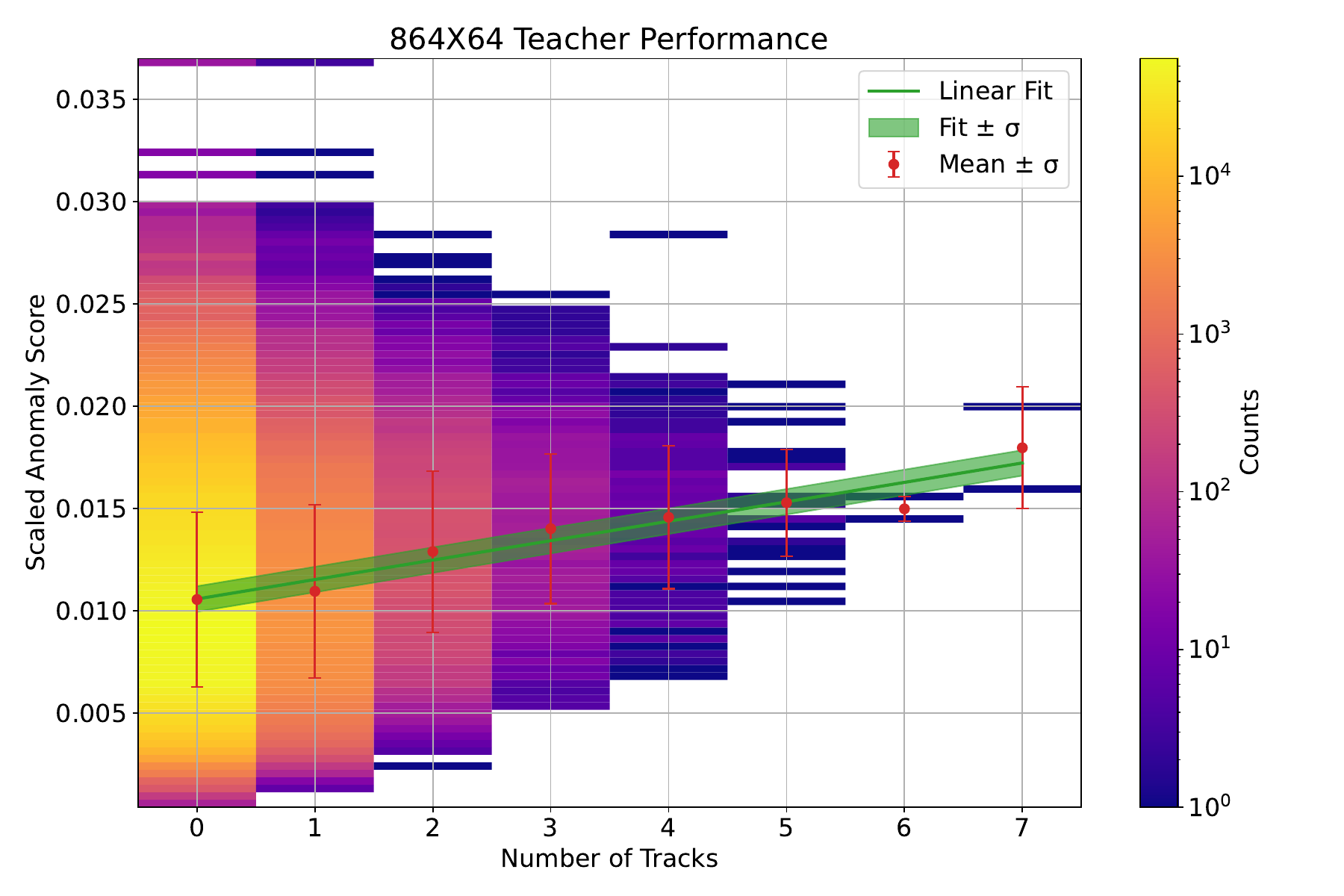}
    \caption{Correlation between scaled Teacher anomaly score (anomaly score divided by total pixel intensity) and number of tracks. The zero-anomaly-scores have been removed from the figure and the fit. The increasing trend suggests that the model is sensitive to event (interaction) topology, not just trivially pixel intensity.}
    \label{fig:normalized_teacher_score}
\end{figure}

\begin{figure}[htbp]
    \centering
    \includegraphics[width=\textwidth]{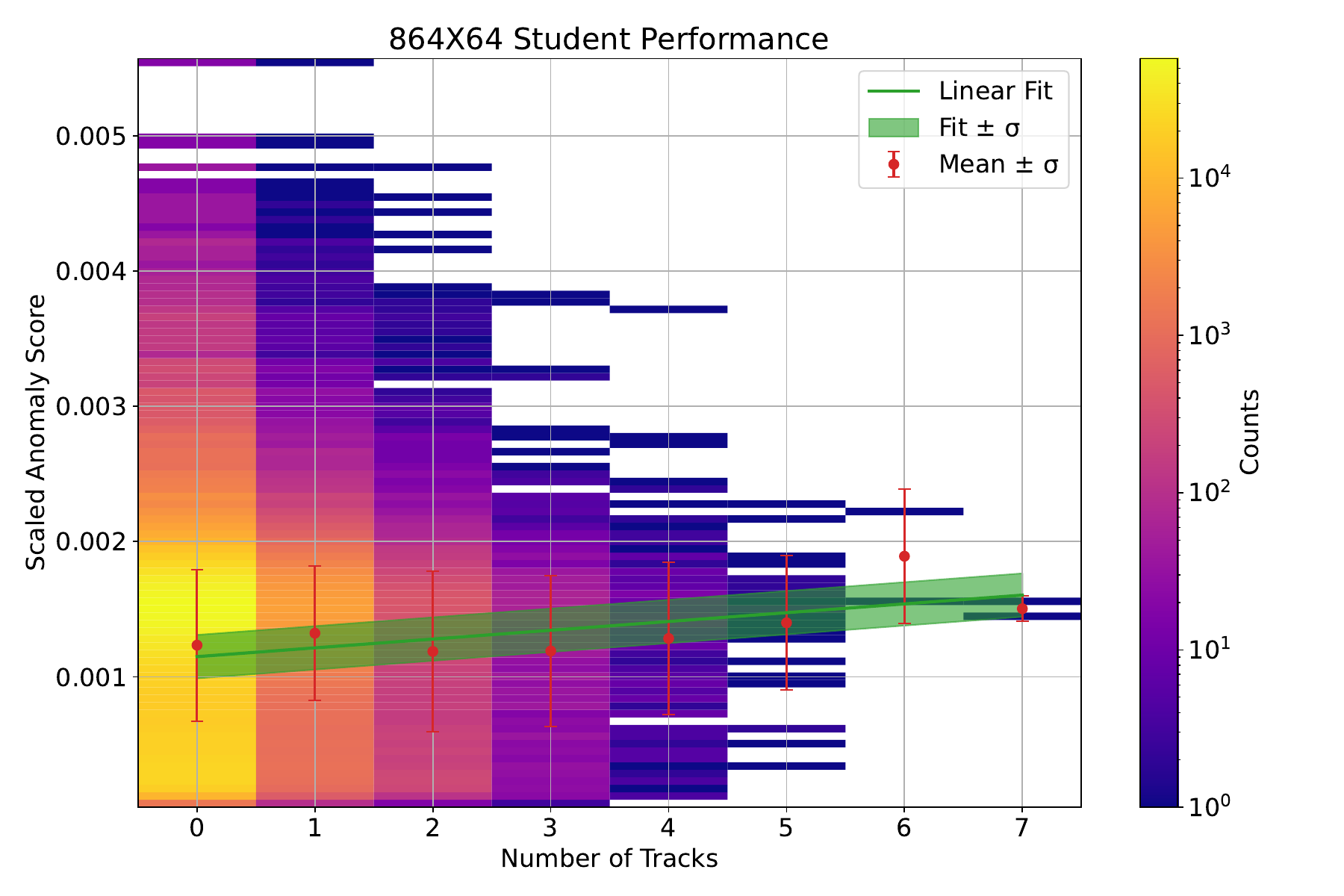}
    \caption{Correlation between scaled Student anomaly score and number of tracks. The zero-anomaly-scores have been removed from the figure and the fit.}
    \label{fig:normalized_student_score}
\end{figure}

Figures~\ref{fig:normalized_teacher_score} and \ref{fig:normalized_student_score} present this scaled correlation. For both Teacher and Student networks, we observe that the scaled anomaly score still increases with the number of tracks, confirming that the networks respond to the topological features of any given interaction contained within a segment, not merely the signal intensity. Zero-anomaly-score bins have been removed in these figures to reduce bias from empty inputs.

To quantify this trend, we performed linear fits of the scaled anomaly score versus $N_{\mathrm{track}}$, summarized in Table~\ref{tab:scaled_score_slopes}. When the y-intercept is constrained to zero, the fitted slopes are positive at greater than $2\sigma$ significance in all cases. This demonstrates that the increase of scaled anomaly score with track multiplicity is statistically robust. The zero-slope test $p$-values for Figure~\ref{fig:normalized_teacher_score} and Figure~\ref{fig:normalized_student_score} are $1.4\times10^{-4}$ and $6.4\times10^{-2}$, respectively, confirming a positive correlation.

\begin{table}[htbp]
\centering
\caption{Linear fit results for the scaled anomaly score as a function of track multiplicity. We report the fitted slope $\beta$, its statistical uncertainty $\sigma_\beta$, and the significance $\beta/\sigma_\beta$. Results are shown for fits with a freely varying ordinate-intercept and with the ordinate-intercept fixed to zero.}
\label{tab:scaled_score_slopes}
\begin{tabular}{cc|ccc}
\hline
 &  & Figure~\ref{fig:raw_score} & Figure~\ref{fig:normalized_teacher_score} & Figure~\ref{fig:normalized_student_score} \\
\hline
\multirow{3}{*}{Free ordinate-intercept} 
 & $\beta$ & $7.7\times10^{-2}$ & $9.5\times10^{-4}$ & $6.5\times10^{-5}$ \\
 & $\sigma_\beta$ & $2.5\times10^{-2}$ & $1.1\times10^{-4}$ & $2.9\times10^{-5}$ \\
 & $\beta/\sigma_\beta$ & 3.1 & 8.6 & 2.2 \\
\hline
\multirow{3}{*}{Fixed ordinate-intercept at 0} 
 & $\beta$ & $6.8\times10^{-2}$ & $3.1\times10^{-3}$ & $3.0\times10^{-4}$ \\
 & $\sigma_\beta$ & $1.3\times10^{-2}$ & $5.3\times10^{-4}$ & $5.9\times10^{-5}$ \\
 & $\beta/\sigma_\beta$ & 5.2 & 5.8 & 5.1 \\
\hline
\end{tabular}
\end{table}

\begin{figure}[htbp]
    \centering
    \includegraphics[width=\textwidth]{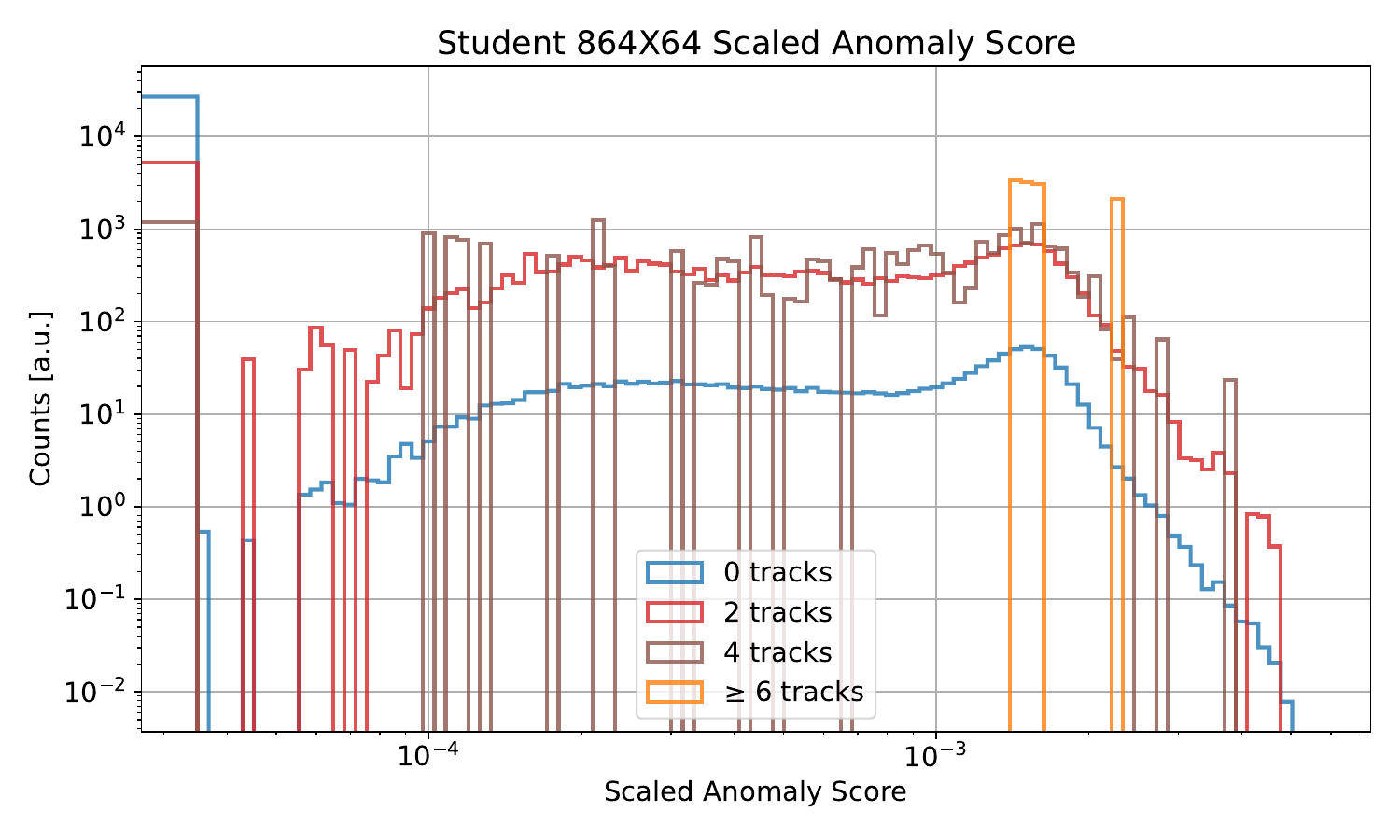}
    \caption{Area-normalized distributions of the scaled Student anomaly score for segments with 0, 2, 4, and $\geq6$ tracks. The separation highlights that the network is truly learning the topology of the input segment.}
    \label{fig:normalized_student_distribution}
\end{figure}

Figure~\ref{fig:normalized_student_distribution} shows the normalized distribution of the scaled Student anomaly score for segments with 0 tracks and for segments with 5 or more tracks. Compared to Figure~\ref{fig:teacher_distribution}, the separation between the two classes is even more pronounced. The 0-track distribution remains sharply localized near zero, while multi-track segments cluster more distinctly around a higher scaled score range. This improved separation confirms that the Student network, trained via knowledge distillation from the Teacher, effectively preserves the discriminative power of the Teacher in isolating multi-track segments, all while producing output in a form that is both scale-invariant and well-suited for fixed-point quantization.

\subsection{Network Performance}
\begin{figure}[htbp]
    \centering
    \includegraphics[width=\textwidth]{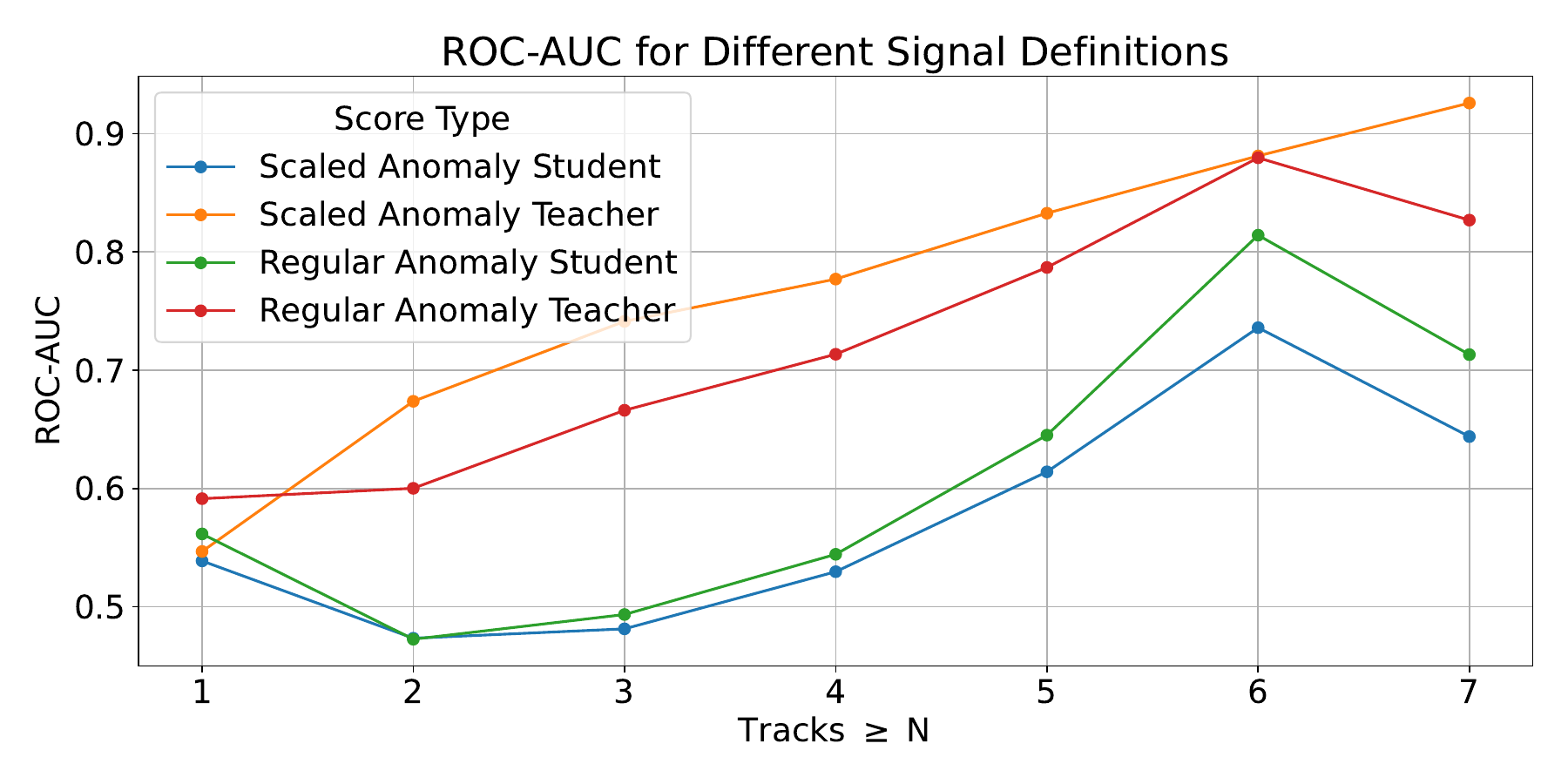}
    \caption{ROC-AUC values for different signal definitions with models trained on $864\times64$-sized input segments. The signal in each entry was defined as having $n$ or more tracks.}
    \label{fig:ROC_AUC}
\end{figure}

To evaluate the practical application of this anomaly detection framework, we quantify its performance by looking at the Receiver Operating Characteristic (ROC) curve, and the area under curve (AUC)~\cite{BRADLEY19971145} values for different signal definitions. The ROC curve is drawn in the false positive rate-true positive rate space, and the AUC value indicates the effectiveness of the signal selection, with 1 being complete separation, and 0.5 being equivalent to random guessing. Figure~\ref{fig:ROC_AUC} shows a summary of the different AUC values for different anomaly score requirements applied to $864\times64$-sized input segments. The target signal is defined as having $n$ or more tracks. Again, the 0-track, 0-anomaly-score inputs were removed beforehand to reduce bias from empty inputs.

As our target hardware application makes use of the Student network, we also report the ROC-AUC values for different Student networks in Table~\ref{tab:ROC_AUC_student}. In all cases, signal segments were defined as those containing three or more truth-level tracks ($n=3$), as the Student network's ability to discriminate improves noticeably beyond this threshold. While the Student achieves strong performance for larger segment sizes, particularly $864\times64$, the ROC-AUC drops to nearly random for $18\times16$. This degradation is expected, as the smallest input window lacks sufficient spatial context to capture global features such as full track structures or multi-track topologies, severely limiting the model’s ability to distinguish signal from background.

Smaller segment sizes, however, are significantly more favorable in terms of hardware resource usage and latency, critical factors for real-time triggering applications. Therefore, a compromise between model performance and deployability must be found, and the optimal segment size will need to be determined through a broader joint optimization of accuracy, resource footprint, and throughput.

\begin{table}[htbp]
\centering
\caption{ROC-AUC values for Student anomaly scores over several segment sizes. The signal in each entry was defined as having $n=3$ or more tracks. \label{tab:ROC_AUC_student}}
\smallskip
\begin{tabular}{c|cc}
\hline
{Segment Size} & {Student} & {Scaled Student} \\
\hline
$864\times64$ & 0.9341 & 0.9335 \\
$64\times32$ & 0.8535 & 0.8527 \\
$18\times16$ & 0.4991 & 0.4998 \\
\hline
\end{tabular}
\end{table}

\subsection{Functional Equivalence of Hardware Implementation}
A key goal of our work is to enable real-time anomaly detection on resource-constrained hardware, including FPGAs, GPUs, or CPUs. To this end, we convert the trained Student network into a hardware-synthesizable version using \texttt{hls4ml}~\cite{Duarte_2018}and simulate its behavior and resource utilization with Xilinx Vivado High-Level Synthesis (HLS) tools~\cite{Xilinx:VivadoHLS}.

\begin{figure}[htbp]
    \centering
    \includegraphics[width=\textwidth]{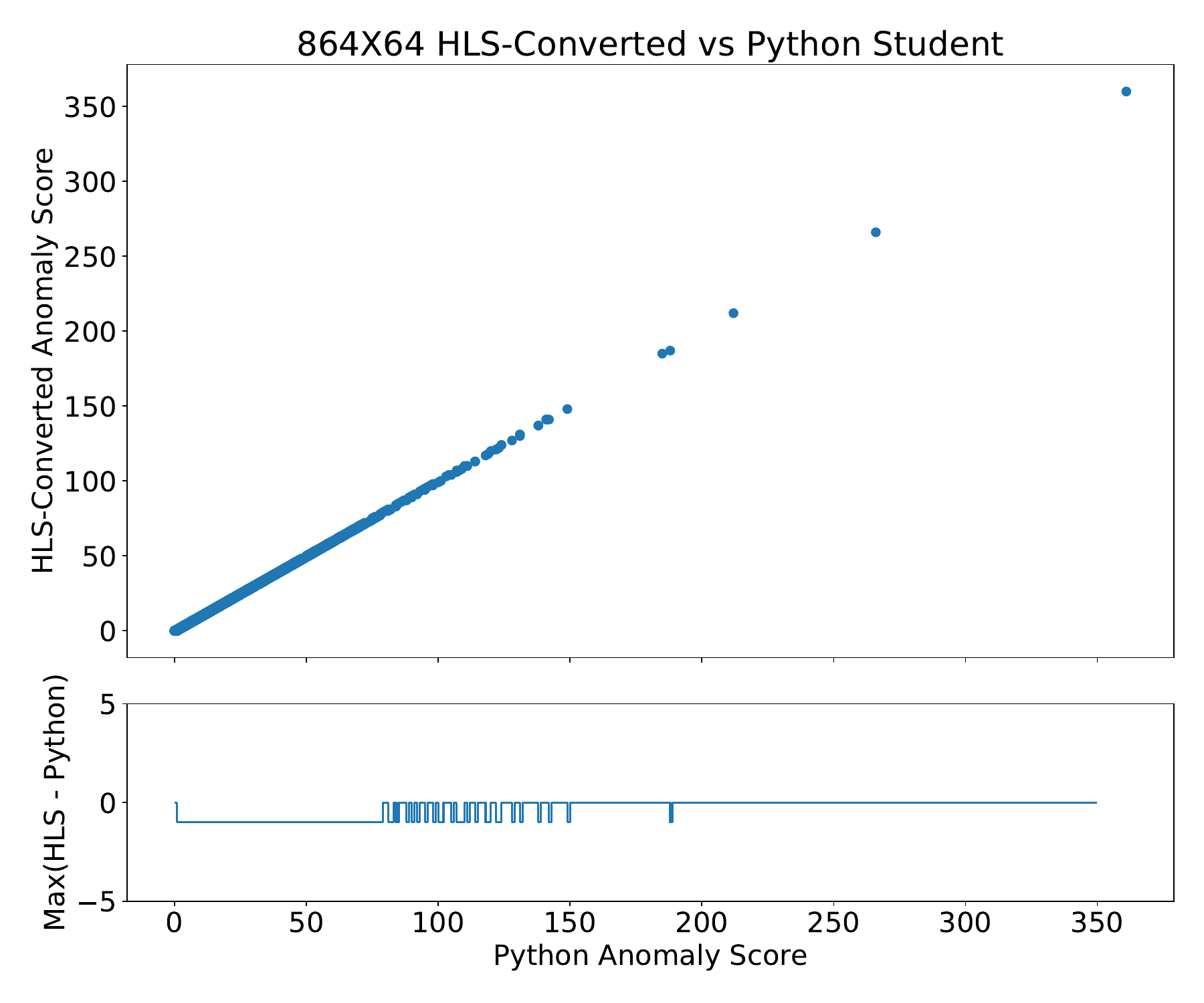}
    \caption{Correlation and maximum difference of anomaly scores from the original Python-based Student model and its hardware-compiled HLS version. The two outputs show agreement (within rounding uncertainty) across all tested inputs, confirming functional equivalence.}
    \label{fig:hls_vs_python}
\end{figure}

To validate the accuracy of the hardware implementation, we compare the anomaly scores generated by the original Python-based Student network using QKeras and its HLS-translated counterpart. Figure~\ref{fig:hls_vs_python} shows the correlation and absolute difference distribution across multiple input segments for the $864\times64$-sized input. The match between scores, being identical or only differing by 1, confirms that the HLS-converted model preserves the essential behavior of the original network, enabling direct deployment onto FPGAs without loss of performance.

\subsubsection{Resource Utilization and Latency}
We also evaluate the expected resource utilization and latency on a Alveo U250 FPGA~\cite{AMD:AlveoU250} of the compiled model using Vivado synthesis reports. Resource metrics include digital signal processing (DSP) blocks, flip-flops (FFs), and look-up tables (LUTs), while timing performance is quantified via estimated clock cycles per inference. Note that the synthesis over-reports LUTs and FFs compared to the actual implementation.

Results for two of the Student models (corresponding to segment sizes $64 \times 32$ and $18 \times 16$) are summarized in Table~\ref{tab:hls_resource_comparison}. Both the total available resources and Super Logic Region (SLR) resources are shown. The SLR is a single FPGA die slice; the Alveo U250 includes 4 SLRs in total. Ideally, the network should fit onto one SLR, not crossing different regions. For each resource, the reported usage is compared to the total available on the device, with the per-SLR capacity given in parentheses; the utilization is defined as the ratio of required to available resources. Note that the model trained for the $864 \times 64$ segment size was not converted due to its large size. Current hardware estimates show that the distilled Student networks satisfy low-latency inference, being below the target clock period of $5~\mathrm{ns}$. However, the models require more than the available resources for the targeted FPGA. Current efforts are ongoing to further reduce network computational footprint and/or targeting different FPGA chips, while maintaining network accuracy and latency performance.

As an alternative to FPGAs, CPU implementation of the models was also explored. We ran inference on a single thread on an Intel Xeon Gold 6226 CPU running at $2.70~\mathrm{GHz}$, and compared the networks' latency to what would be required for MicroBooNE, namely $3.2~\mathrm{ms/readout}$ \cite{MicroBooNE:2015bmn}. Table~\ref{tab:CPU_resource_comparison} shows the latency for the different Student models, per input segment and readout, where each readout is an image of $3456\times640$-pixel size. We also present the number of threads needed to process one readout within the required latency in order to keep up with the expected raw data rate for the experiment. This allows us to estimate the requirements for parallelization. As one Xeon Gold 6226 CPU has 12 cores with 24 threads, our results indicate that a single CPU can host enough parallel instances of the same model, one per input segment, to keep up with the data stream.

This metric is intended to provide a rough estimate of the hardware requirements, demonstrating that the system can process the data stream in real time without requiring an unmanageable number of CPUs. Even when accounting for hyperthreading efficiency and potential memory bus contention, the total compute required remains on the order of $O(1)$ CPUs, which is feasible for our setup.

\begin{table}[htbp]
\centering
\caption{Comparison of resource usage and estimated timing for different segment sizes ran on FPGA simulation. The target clock period is $5~\mathrm{ns}$. For each resource (DSP, FF, and LUT), the first number shows the estimated usage of the network, followed by the total available on the device and the per-SLR capacity in parentheses. The utilization is defined as the ratio of the required usage to the available resource.}
\label{tab:hls_resource_comparison}
\begin{tabular}{c|cc}
\hline
{Segment Size} & {64$\times$32} & {18$\times$16} \\
\hline
Estimated Clock Period (ns) & 4.543 & 4.419 \\
\hline
{DSP Usage} & 26872 / 12288 (3072) & 5329 / 12288 (3072) \\
Utilization (\%) & 218\% (SLR: 874\%) & 43\% (SLR: 173\%) \\
{FF Usage} & 918724 / 3456000 (864000) & 146143 / 3456000 (864000) \\
Utilization (\%) & 26\% (SLR: 106\%) & 4\% (SLR: 16\%) \\
{LUT Usage} & 2202970 / 1728000 (432000) & 305372 / 1728000 (432000) \\
Utilization (\%) & 127\% (SLR: 509\%) & 17\% (SLR: 70\%) \\
\hline
\end{tabular}
\end{table}

\begin{table}[htbp]
\centering
\caption{Latency and parallelization requirements for different segment sizes for network ran on CPU.}
\label{tab:CPU_resource_comparison}
\begin{tabular}{c|ccc}
\hline
{Metric} & {864$\times$64} & {64$\times$32} & {18$\times$16} \\
\hline
Latency / Segment (ms) & 1.911 & 0.0594 & 0.0090 \\
Latency / Readout (ms) & 76.44 & 64.13 & 69.32 \\
Threads Needed & 23.89 & 20.04 & 21.66 \\
\hline
\end{tabular}
\end{table}

\section{Discussion}
\label{sec:discussion}

In this work, we developed and studied an anomaly detection framework for LArTPCs based on convolutional autoencoders and knowledge distillation. Using the MicroBooNE Open Dataset as a benchmark, we demonstrated that the Teacher network originally developed for anomaly detection applications in CMS can effectively identify rare interactions in LArTPCs, particularly those with multi-track topology.

To enable deployment on resource-constrained hardware platforms, we employed knowledge distillation to train a lightweight Student network in performing comparably to the Teacher network. The supervised model approximates the Teacher's output while being significantly more computationally efficient, allowing implementation on CPUs while satisfying the latency and data throughput requirements of experiments such as MicroBooNE and SBND.

Our results show that both raw and scaled anomaly scores correlate with physical event (interaction) complexity, specifically the number of ionizing tracks in any given input segment. The Student network reproduces this behavior while remaining compatible with online execution constraints, making it suitable for integration into LArTPC data acquisition and trigger systems.

Together, these findings highlight that knowledge distillation successfully transfers the discriminating capability of the Teacher to the Student model, preserving sensitivity to multi-track topologies while ensuring the compactness and efficiency needed for deployment. By producing scale-invariant outputs that are naturally compatible with fixed-point quantization, the Student model is directly amenable to FPGA-based implementation.

In practice, this means that the scaled anomaly score produced by the Student network can serve as a trigger criterion, enabling low-latency identification and selection of complex, topologically rich interactions. Such application opens a path toward intelligent, AI-assisted data selection in future LArTPC experiments, with the potential to maximize physics reach in a model-agnostic way. Alternatively, this anomaly detection scheme can serve as a ``filter'' for data down-selection or localization of high-multiplicity regions of interest.

\section{Summary and Outlook}
\label{sec:summary}

This study marks an initial step toward viable hardware implementations of real-time  anomaly detection for LArTPCs using neural networks. We applied knowledge distillation to reduce a high-capacity Teacher model into an efficient Student network, capable of operating on CPUs, and potentially also FPGAs with additional model downsizing. Using real and simulated LArTPC data, we demonstrated the model's ability to identify complex event topologies within the data and confirmed its potential through performance benchmarks and resource synthesis studies.

Future work will focus on advancing several aspects of this research. On the technical side, continued effort will be devoted to optimizing model architectures with respect to latency, power consumption, and resource utilization efficiency, with the long-term goal of implementing these networks on increasingly power-aware platforms such as FPGAs. From a physics perspective, validation on rare SM or BSM physics signatures will be essential to assess the applicability of the approach to existing or planned experiments. Equally important will be the integration of the Student model into real-time DAQ systems, enabling online testing within operating LArTPC experiments with realistic detector conditions. Finally, attention should be given to robustness, particularly through domain adaptation techniques that can ensure generalization across different detector geometries and noise environments.

Our current setup used a relatively simple Teacher architecture and training procedure. More sophisticated strategies exist to improve the performance of both Teacher and Student models, such as outlier exposure~\cite{hendrycks2019deepanomalydetectionoutlier}, co-learning~\cite{pol2023kd}, or other advanced training techniques. Exploring these methods could further enhance the network's ability to capture anomalous features in LArTPC data.

These efforts aim to contribute to the development of intelligent triggers for neutrino detectors and open new possibilities for AI-driven low-level data selection.


\acknowledgments
This work has been supported by the National Science Foundation under OAC-2209917. The authors thank Giuseppe Cerati for valuable help with utilizing the MicroBooNE Open Dataset. 


\bibliographystyle{JHEP}
\bibliography{biblio.bib}

\end{document}